\documentclass[aps,pre,twocolumn]{revtex4}

\newcommand{\be}{\begin{equation}}
\newcommand{\ee}{\end{equation}}
\newcommand{\ba}{\begin{eqnarray}}
\newcommand{\ea}{\end{eqnarray}}
\newcommand{\bc}{\begin{center}}
\newcommand{\ec}{\end{center}}
\newcommand{\bi}{\begin{itemize}}
\newcommand{\ei}{\end{itemize}}
\newcommand{\ie}{{\it i.e. }}

\newcommand{\ignore}[1]{}

\begin{document}

\title{
Weak Wave Turbulence Scaling Theory for Diffusion and
Relative Diffusion in Turbulent Surface Waves
}
\author{V\'{\i}ctor M. Egu\'{\i}luz}
\author{Mogens T. Levinsen}
\author{Preben Alstr\o m}
\affiliation{
Center for Chaos and Turbulence Studies, The Niels Bohr Institute,
Blegdamsvej 17, DK-2100 Copenhagen \O, Denmark  
}

\begin{abstract}
We examine the applicability of the weak wave turbulence theory in explaining 
experimental scaling results obtained for the diffusion and relative diffusion
of particles moving on turbulent surface waves.
For capillary waves our
theoretical results are shown to be in good agreement with experimental
results, where a distinct crossover in diffusive behavior is observed at the
driving frequency. For gravity waves  our results are discussed in the light of
ocean wave studies. 
\end{abstract}

\maketitle

\section{Introduction}

The study of particles moving on surface waves has shown that the particle
motion often is far from being Brownian
\cite{Ramshankar90,Ramshankar91,Schroder97a,Schroder97b,Hansen97}. Similar
results are found in ocean studies \cite{Okubo71}. The turbulence
observed in surface waves is  strongly influenced by the presence of a
dispersion relation (for deep-water surface waves) $\omega_k^2 = gk+\sigma
k^3/\rho$ between the wave oscillation frequency and the wave vector $k$
\cite{dispersion}. Here $g$ is the gravitational acceleration, $\sigma$ is the
surface tension coefficient, and $\rho$ is the density. Experimentally surface
waves can be studied by vertically oscillating a fluid with a free surface. On
the surface of the fluid, waves with a clearly discernible wavelength $\Lambda$
are then formed if the amplitude of oscillations exceeds a critical value, the
so-called Faraday instability \cite{nevolin}. The excitation of surface waves
in the Faraday system is a parametric effect giving rise to a fundamental wave
frequency $\Omega$ equal half the driving frequency. The frequency $\Omega$ and
wavelength $\Lambda$ are associated through the dispersion relation. At small
frequencies, $k<k_a=(g\rho/\sigma)^{1/2}$, the first term in the dispersion
relation dominates, and the waves formed are called gravity waves (typical of
ocean waves). At large frequencies ($k>k_a$) the effect of gravity can be
neglected, and the second term dominates. The waves thus formed are called
capillary waves. For water, the wave length $a=2\pi/k_a$, separating gravity
and capillary waves, is about 2 cm.

Several studies have been carried out to measure the diffusivity of particles
moving on capillary surface waves generated above the Faraday instability
\cite{Ramshankar90,Ramshankar91,Schroder97a,Schroder97b,Hansen97,Schroder96}.
Two particles separated by a distance $R$ move more rapidly apart as $R$ is
increased. The relative diffusivity can be defined as
\be
\langle \frac{dR^2}{dt} \rangle = 2 R \langle \delta v \rangle~,
\ee
where $\delta v$ is the longitudinal component of the velocity difference. The
variance $\sigma^2$ of the relative diffusivity $dR^2/dt$ is related to the
velocity difference squared,
\ba
\left( \frac{\sigma}{R}\right)^2 &\sim& \langle |\delta v(R)|^2\rangle~.
\label{sigv}
\ea
It has been found experimentally that in capillary waves $\left(
\frac{\sigma}{R}\right)^2 \propto 1 - a (R/\Lambda)^{-1/4}$, with $a=0.66$
\cite{Schroder96}.

The purpose of the present investigation is to examine the applicability and 
limitations of the weak wave turbulence theory in explaining 
experimental scaling results obtained for the diffusion and relative diffusion
of particles moving on turbulent surface waves.
Diffusion theories based on random wave fields have
likewise been considered \cite{RWF}, and Herterich and
Hasselmann \cite{Herterich}  particularly considered diffusion by
surface gravity waves. For relative 
diffusion on capillary waves we
find the exponent $-1/4$ as in experiments in the regime $R > \Lambda$. For
gravity waves, we find an exponent $7/6 \sim 1.14$, a result which is
intriguingly close to the value $1.15$ found in ocean studies by Okubo
\cite{Okubo71}.
We note that the
relative diffusion exponent 4/3 obtained in fully
developed turbulence theory previously have been
used in the discussion of experimental data \cite{Monin75}.

To obtain 
the diffusivity, the particle displacement $\Delta x (t)=x(t_0+t)-x(t_0)$ over
a time $t$ is measured along an arbitrary axis for many initiation times $t_0$.
From the resulting distribution the variance $V(t)=\langle [\Delta
x(t)]^2\rangle$ is found, and the diffusivity $D$ can be extracted, $V(t)=2Dt$.
For Brownian motion the diffusivity $D$ is a constant. For deep water capillary
waves (we only consider deep-water surface waves)
formed at high frequencies, the diffusion has been found
experimentally to be anomalous
\be
V(t) \sim t^\lambda~,
\ee
with $\lambda \geq 1$ (super-diffusive)
\cite{Schroder97a,Schroder97b,Hansen97,Schroder96}. The exponent is observed to
change drastically from a value $\lambda = 1.6 - 1.9$ at length scales below
$\Lambda$ to a value $\lambda = 1.0 - 1.3$ at length scales above $\Lambda$.
The values of $\lambda$ is observed  to decrease with increasing drive; at
large drive $\lambda$ attains the Brownian motion value
1\cite{Schroder97a,Schroder97b}. Assuming a connection between spatial and
temporal correlations we find for capillary waves $\lambda = 16/9$ for length
scales below $\Lambda$, and $\lambda =1$ for length scales above, \ie values rather
close to the experimental values. Ocean studies have been performed using
floaters \cite{Stommel49}, chemical tracers \cite{Okubo71}, and near surface
drifters. To what extent these studies relate to surface waves is unclear. Our
results do, however, indicate a connection at least for the chemical tracers,
where Okubo finds an exponent of 2.34
and from weak wave turbulence theory we find $8/3$.

\section{Spatial correlations and relative diffusion}

Spatial correlations between pairs of particles at ${\bf r}$ and ${\bf r} +
{\bf R}$ are related to the velocity difference squared
\ba
\langle |\delta v(R)|^2\rangle
&=& \langle |{\bf v} ({\bf r}+ {\bf R}) - {\bf v}({\bf r})|^2\rangle
\nonumber \\
&=& 2[\langle |{\bf v}({\bf r})|^2\rangle - \langle {\bf v} ({\bf r} + {\bf R})
\cdot {\bf v} (\bf r) \rangle ]~.
\ignore{
&\sim& \int k \omega_k n_k g(k/k_F) (1-e^{i kR\cos\theta}) d{\bf k}~\\
&\sim&\int_0^{\infty} k^{2+\alpha-\beta} g(k/K) [1-J_0(kR)] dk~,
}
\label{vds}
\ea
Following the
calculations of Ref.~\cite{Schroder98}, we have an expression for the
space-correlation of the velocity
\ba
{\cal C} (R) = \langle {\bf v}( {\bf r}) {\bf v} ({\bf r}+{\bf R}) \rangle
&\sim& \int k \omega_k n_k e^{i kR\cos\theta} d{\bf k}
\nonumber \\
&\sim& \int_0^{\infty} k^2 \omega_k n_k J_0(kR) dk~,
\label{spacecor}
\ea
where $n_k$ is the isotropic Kolmogorov spectrum \cite{Zakharov92}, and
$J_0$ is the zeroth order Bessel function.


For capillary waves, the dispersion relation is $\omega_k=(\sigma/\rho)^{1/2}
k^{\alpha}$ with $\alpha=3/2$, and the isotropic Kolmogorov spectrum (three
wave interactions) is \cite{Zakharov92,Zakharov67}
\be
n_k\sim P^{1/2}\rho^{3/4}\sigma^{-1/4}k^{-\beta}~,
\ee
where $\beta =17/4$ and $P$ is the energy flux assumed to be constant.
From measurements of the spectrum of wave-amplitudes \cite{henrya,henryb} 
we know that
the spectrum levels off below $K=2\pi/\Lambda$. In order to account for this
change we introduce a scaling function $g(k/K)$ across the drive multiplying
$n_k$. At $k<K$, $g(k/K)$ must rise from  zero in such a way as to quench the
decrease of $n_k$. At $k>K$,  $g(k/K)$ can be assumed constant (up to the
inverse dissipative scale $k_d$ where the theory breaks down).
The motion at length
scales below the dissipative scale is expected to become ballistic. Also the
system size L is a relevant length scale. Obviously the diffusion change
character near the system size since particles cannot travel longer 
distances than L. Also we shall consider the characteristic length 
scale $a$ separating capillary from gravity waves.

The integral for the spatial correlations is given by
\be
{\cal C} (R) \sim \int k^{2+\alpha- \beta} g(k/k_F) J_0(kR) d{\bf k}~
\ee
where the specific $k$ dependence for $\omega_k$ and $n_k$ has been inserted.
For capillary waves $2+\alpha- \beta = -3/4$.
This integral depends explicitly on $R$. Consider the value $K_R\simeq
2.4/R$, the lowest $k$ value at which $J_0(kR)=0$. For $K_R > K$, the dominant
contribution to the integral comes from the range $K<k<K_R$, where the scaling
function $g$ is constant,
\ba
{\cal C} (R) &\sim& \int_0^{\infty} k^{2+\alpha-\beta} g(k/K) J_0(kR) dk
\nonumber \\
          &\sim& k^{3+\alpha-\beta}|^{K_R}_{K} \sim R^{-3-\alpha+\beta}.
\label{scal}
\ea
The result for the velocity difference is
\be
\langle |\delta v(R)|^2\rangle \sim 1-b(KR)^{-3-\alpha+\beta} \sim
1-b(KR)^{-1/4}~,
\label{dv1}
\ee
where $b$ is a constant of order one \cite{note}). 

For $K_R < K$, the Bessel  function oscillations set in below $K$ while the
integrand is still increasing in size due to the behavior of the scaling 
function $g(k/K)$. Therefore the dominant contribution to the integral comes
from a peak centered around $K$, it is oscillatory in nature
with a vanishing envelope falling off like $R^{3/2}$. 
For large $R$ values we essentially have $\langle |\delta v(R)|^2\rangle$
constant.


For gravity waves the dispersion relation is $\omega=g^{1/2}k^{\alpha}$ with
$\alpha=1/2$ (four-wave interaction), and we have energy flow towards higher
frequencies compared to $\Omega$ giving rise to the isotropic Kolmogorov
spectrum \cite{Zakharov92,Zakharov66}
\be
n_k\sim P^{1/3}\rho^{2/3}k^{-\beta}~,
\ee
where $\beta =4$ and $P$ is the energy flux assumed to be constant. Below
$\Omega$ a constant wave-number flux $Q$ towards lower frequencies yields
\cite{Zakharov82}
\be
n_k\sim Q^{1/3}\rho^{2/3}g^{1/6}k^{-\beta+\alpha/3}~,
\ee
with $\beta-\alpha/3=23/6$.

As above we invoke a scaling function $g(k/K)$ describing the change in the
behavior of $n_k$ as $k$ crosses the wavenumber $K$ of the drive. According to
the form of the spectra we have
\be
g(k/k_F) \sim \left\{\begin{array}{ll}
                    1 &\mbox{if $K < k < k_a,k_d$}\\
                    k^{\alpha/3}  &\mbox{if $k_L < k < K$}.
                 \end{array}
\right.
\ee
Other corrections are relevant at length scales outside the above regime.
The motion at sufficiently large length scales ($k<k_L$) may
be expected to become Brownian, as also found in studies of motion of drifters
near the surface of the oceans \cite{Okubo71}.

For the velocity difference squared we have
\ba
\langle |\delta v(R)|^2\rangle
&\sim& \int_0^{\infty} k^{2+\alpha-\beta} g(k/K) [1-J_0(kR)] dk
\nonumber \\
&\sim& \int_{K_R}^{\infty} k^{-3/2} g(k/K) dk~.
\ea

For $K_R > K$, the scaling function is constant in the integration
region
above $K_R$, and we have
\be
\langle |\delta v(R)|^2\rangle \sim K_R^{-1/2} \sim (KR)^{1/2}~.
\label{gr1}
\ee
For $K_R < K$ the part of the integration, where $g(k/K)\sim k^{1/6}$,
gives the $R$ dependent contribution, and
\be
\langle |\delta v(R)|^2\rangle + const \sim K_R^{-1/3}\sim
(KR)^{1/3}~.
\label{gr2}
\ee
The growth of $\langle |\delta v(R)|^2\rangle$ stops at length scale
$L$.
We note that the exponent 1/3 for the velocity difference squared implies
an exponent 7/6 for the relative diffusivity measure $\sigma$. It is
interesting that Okubo \cite{Okubo71} finds the relative diffusion exponent 
to be 1.15 for oceanic diffusion, although we should emphasize that the 
relative diffusion 
exponent 4/3 obtained in fully developed turbulence theory has generally 
been applied to explain experimental data \cite{Monin75}. 

\section{Temporal correlations and capillary waves}

Next we consider single particle diffusion in capillary waves.
For arbitrary times, the mean-square displacement $\langle r^2 \rangle=\langle
|r(t_0+t)-r(t_0)|^2 \rangle$ is expressed in terms of the Lagrangian velocity
correlation function  ${\cal C} (\tau) = \langle {\bf v} ({\bf r} (t)) {\bf v}
({\bf r}(t+ \tau)) \rangle$ as \cite{hansen}
\be
\langle r^2 \rangle = 2 t \int_0^t (1 - \tau/t) {\cal C} (\tau) d \tau~.
\label{displ}
\ee
In the Taylor limit of very small $t$, ${\cal C}(\tau) \simeq \langle v_L^2
\rangle$, where $\langle v_L^2\rangle$ is the mean-square Lagrangian velocity,
to be determined by sampling along particle orbits. In this case
\be
\langle r^2 \rangle \approx \langle v_L^2\rangle t^2 ~.
\label{ball}
\ee
For incompressible flows, the mean-square velocity obtained by Eulerian
sampling, $\langle v_E^2\rangle$, is identical to the mean-square velocity,
$\langle v_L^2\rangle$, obtained from Lagrangian measurements
\cite{Tennekes72}. The subscripts on $\langle v^2\rangle$ will be omitted from
here on. For large times, $t \gg \tau_L$, where $\tau_L = \frac{1}{\langle
v^2\rangle} \int_0^\infty {\cal C}(\tau) d \tau$ is the Lagrangian integral
time scale, Eq.~(\ref{displ}) gives a diffusion-like dispersion
\be
\langle r^2 \rangle \sim 2 \langle v^2\rangle \tau_L t ~.
\label{diff}
\ee

First we consider capillary waves.
If we invoke a scaling form $\langle r^2 \rangle\sim t^\lambda$ at $t<T$, then
Eq.~(\ref{displ})
gives
\be
{\cal C}(\tau)\sim\tau^{\lambda-2}~.
\ee
On the other hand, invoking the same scaling $\langle r^2 \rangle\sim t^\lambda$
and using Eq.~(\ref{scal}) yields
\be
{\cal C}(\tau)\sim(\tau^{\frac{\lambda}{2}})^{-\frac{1}{4}}~.
\ee
The result is a self-consistency condition for the exponent $\lambda$. We find
\be
\lambda = \frac{4}{5+\alpha-\beta} = \frac{16}{9}~.
\label{self}
\ee
This value obtained for $\lambda$ is in
good agreement with experimental observations which vary $\pm$10\% from this
value \cite{Schroder97a,Schroder97b}.
From this the exponent for the time correlation  function becomes $-2/9$.

Above it has tacitly been assumed that the time scale $t$ is larger than the
one corresponding to the dissipative frequency $\omega_d$ corresponding to the
inverse dissipative scale $k_d$. Experimentally, the sampled diffusion
distances at small times may approach the dissipative scale at low values of
the drive. Below this scale $\lambda$ has the ballistic value 2, and the
observed slope (1.8--1.9) slightly larger than the theoretical value 16/9 may
be an effect hereof.

We have also assumed that the particles can follow the fluid velocity field,
i.e. if ${\bf v}$ is the velocity field of the particles and ${\bf u}$ is that
of the fluid, our assumption is that ${\bf v}={\bf u}$, or at least that the
weak ansatz ${\cal C}(\tau) = \langle {\bf u}(t+\tau){\bf u}(t)\rangle$ is
fulfilled. However, at large driving amplitudes the particles may not follow
the fluid velocity. As an example, assume that corrections for the velocity
correlation takes the following form
\be
\langle {\bf v}(t){\bf v}(t+\tau) \rangle \sim \tau^{-\delta}
\langle {\bf u}(t){\bf u}(t+\tau) \rangle
\ee
where ${\bf v}$ and ${\bf u}$ are the velocity fields of the particles and of
the flow respectively. Following the calculation above, we then find
\be
\lambda = \frac{4-2\delta}{5+\alpha-\beta} = \frac{16-8\delta}{9}~,
\ee
which may explain why a slightly smaller exponent than 16/9 is observed for
higher driving amplitudes (e.g. $\delta=0.2$ gives an exponent $\lambda=1.6$).

Equivalently, the self-consistent condition for $\lambda$ (Eq.~(\ref{self})),
can be obtained  in terms of the energy spectrum
\be
E_k \sim k^{-\nu}~,
\ee
where the energy per area is
\be
E=\int E_{\bf k} d{\bf k}= \int E_k dk = \int k \omega_k n_k dk~,
\label{trans}
\ee
we have the relationship $\nu = -(1+\alpha - \beta)$, and
\be
\lambda= \frac{4}{4-\nu}~.
\label{lam}
\ee
The capillary wave exponent is $\nu = 7/4$.

For $t>T$, the time
correlation ${\cal C} (\tau)$ is oscillatory and therefore negligible for $t>T$.
In this case, Eq.~(\ref{diff}) applies
\be
\langle r^2 \rangle \sim t~,
\ee
and $\lambda=1$. The Brownian motion value obtained for $\lambda$ at times
$\tau$ above $T$ is in agreement with experimental observations for large
drives.
The slightly larger values of $\lambda$ at lower drives may
arise from corrections to scaling. Experimentally, at sufficiently large times
(above those considered in \cite{Schroder97a,Schroder97b}),
the sampled diffusion distances
are limited by the systems size $L$ and eventually one must have $\lambda=0$.

\section{Temporal correlations and gravity waves}

For gravity waves, we can follow a similar assumption as the one proposed for
capillary waves. However, in contrast to the situation for capillary waves, the
scaling is given for $\langle |{\bf v} (t) - {\bf v} (t+ \tau)|^2 \rangle$. Thus,
for $\tau < T$, we have from Eq.~(\ref{gr1})
\be
\langle |{\bf v} (t) - {\bf v} (t+ \tau)|^2 \rangle \sim \tau^{\lambda/4}~,
\ee
and for $\tau > T$, we get from Eq.~(\ref{gr2})
\be
\langle |{\bf v} (t) - {\bf v} (t+ \tau)|^2 \rangle \sim \tau^{\lambda/6}~.
\ee
Assuming
\be
\sigma \sim \frac{R^2}{t}~,
\ee
the exponents can now be calculated using Eq.~(\ref{sigv})
from the self-consistent condition
\be
t^{\lambda-1}\sim \left\{
\begin{array}{ll}
t^{5\lambda/8} \ &\mbox{small $t$}	\\
t^{7\lambda/12} \ &\mbox{large $t$} 	
\end{array}
\right.
\ee
The values obtained for $\lambda$ are $8/3$ for small $t$ and $12/5$ for large
$t$. Interestingly, a similar situation is observed in ocean
studies where $\lambda = 2.34$  is found in several dye studies collected and
analyzed by Okubo \cite{Okubo71}.

\section{Conclusions}

Good agreement is found between theoretical and experimental scaling results 
for the diffusion and relative diffusion of particles on weakly
turbulent surface waves. 
In the case  of relative diffusion, for capillary waves the exponent -1/4 for
the space correlation function is recovered. 
For gravity waves, a scaling behavior is derived for the velocity difference
squared  with the  exponent changing from 1/2 at small distances to 1/3 at
large distances.  This corresponds respectively to scaling exponents 5/4 and
7/6 for the  relative diffusivity. Intriguingly, Okubo finds a relative
diffusion exponent of 1.15 for oceanic diffusion \cite{Okubo71}.
This differs from the
relative diffusion exponent 4/3 obtained in fully
developed turbulence theory, previously used in the
discussion of experimental data \cite{Monin75}.

For capillary waves the theory yields a crossover in the diffusivity at the
wavelength $\Lambda$ with a  change in diffusion exponent $\lambda$ from
$\lambda= 16/9\simeq 1.78$  to $\lambda = 1$, in good agreement with
experimental findings. Particles suspended on the fluid surface may not follow
the fluid far above the Faraday instability, and we suggest that this may
explain the small discrepancy between weak turbulence theory and experimental
results in this regime.  For gravity waves the weak turbulence theory gives a
diffusion exponent of $12/5 = 2.4$ for long times which compares favoritely
with the exponent $2.34$ found by Okubo for dye diffusion. To what extent the
weak wave turbulence theory really explains the oceanic diffusivity is unclear
but it is fascinating to speculate that wave motion taking place mostly on a
short scale could be responsible for the long time diffusion in oceans.
At very long time scales, i.e. above the Lagrangian time scale, ordinary 
Brownian motion is expected.

\section{Acknowledgments}
We acknowledge financial support from the Danish Natural Sciences Research
Council and CORE.


\begin{thebibliography}{}
\bibitem{note} We note that a similar result was derived in 
ref.~\cite{Schroder98}, however, using incorrect limits on the
integral.
\end{thebibliography}
\end{document}